\documentclass[a4paper,fleqn,usenatbib]{mnras}

\usepackage[T1]{fontenc}
\usepackage{ae,aecompl}

\usepackage{graphicx}	
\usepackage{amsmath}	
\usepackage{amssymb}	

\title[The beamed jet and quasar core of 4C~71.07]{The beamed jet and quasar core of the distant blazar 4C~71.07}
\author[C. M. Raiteri et al.] 
{C.~M.~Raiteri              $^{ 1}$\thanks{E-mail:raiteri@oato.inaf.it},
M.~Villata                  $^{ 1}$,
M.~I.~Carnerero             $^{ 1}$,
J.~A.~Acosta-Pulido         $^{ 2,3}$,
\newauthor
D.~O.~Mirzaqulov            $^{ 4}$,
V.~M.~Larionov              $^{ 5,6}$,
P.~Romano                   $^{ 7}$,
S.~Vercellone               $^{ 7}$,
I.~Agudo                    $^{ 8}$,
\newauthor
A.~A.~Arkharov              $^{ 6}$,
U.~Bach                     $^{ 9}$,
R.~Bachev                   $^{10}$,
S.~Baitieri                 $^{11}$,
G.~A.~Borman                $^{12}$,
\newauthor
W.~Boschin                  $^{13,2,3}$,
V.~Bozhilov                 $^{14}$,
M.~S.~Butuzova              $^{12}$,
P.~Calcidese                $^{15}$,
D.~Carosati                 $^{16,13}$,
\newauthor
C.~Casadio                  $^{ 9}$,
W.-P.~Chen                  $^{17}$,
G.~Damljanovic              $^{18}$,
A.~Di~Paola                 $^{19}$,
\newauthor
V.~T.~Doroshenko            $^{20}$\thanks{Passed away},
N.~V.~Efimova               $^{ 6}$,
Sh.~A.~Ehgamberdiev         $^{ 4}$,
M.~Giroletti                $^{21}$,
\newauthor
J.~L.~G\'omez               $^{ 8}$,
T.~S.~Grishina              $^{ 5}$,
S.~Ibryamov                 $^{22}$,
H.~Jermak                   $^{23}$,
S.~G.~Jorstad               $^{24,5}$,
\newauthor
G.~N.~Kimeridze             $^{25}$,
S.~A.~Klimanov              $^{ 6}$,
E.~N.~Kopatskaya            $^{ 5}$,
O.~M.~Kurtanidze            $^{25,26}$,
\newauthor
S.~O.~Kurtanidze            $^{25}$,
A.~L\"ahteenm\"aki          $^{27,28}$,
E.~G.~Larionova             $^{ 5}$,
A.~P.~Marscher              $^{24}$,
\newauthor
B.~Mihov                    $^{10}$,
M.~Minev                    $^{14}$,
S.~N.~Molina                $^{ 8}$,
J.~W.~Moody                 $^{29}$,
D.~A.~Morozova              $^{ 5}$,
\newauthor
S.~V.~Nazarov               $^{12}$,
A.~A.~Nikiforova            $^{ 5}$,
M.~G.~Nikolashvili          $^{25}$,
E.~Ovcharov                 $^{14}$,
\newauthor
S.~Peneva                   $^{10}$,
S.~Righini                  $^{21}$,
N.~Rizzi                    $^{30}$,
A.~C.~Sadun                 $^{31}$,
M.~R.~Samal                 $^{17}$,
\newauthor
S.~S.~Savchenko             $^{ 5}$,
E.~Semkov                   $^{10}$,
L.~A.~Sigua                 $^{25}$,
L.~Slavcheva-Mihova         $^{10}$,
\newauthor
I.~A.~Steele                $^{23}$,
A.~Strigachev               $^{10}$,
M.~Tornikoski               $^{27}$,
Yu.~V.~Troitskaya           $^{ 5}$,
\newauthor
I.~S.~Troitsky              $^{ 5}$,
and O.~Vince                 $^{18}$\\
}
\date{Accepted XXX. Received YYY; in original form ZZZ}

\pubyear{2019}

\begin{document}
\label{firstpage}
\pagerange{\pageref{firstpage}--\pageref{lastpage}}
\maketitle

\begin{abstract}
The object 4C~71.07 is a high-redshift blazar whose spectral energy distribution shows a prominent big blue bump and a strong Compton dominance. We present the results of a two-year multiwavelength campaign led by the Whole Earth Blazar Telescope (WEBT) to study both the quasar core and the beamed jet of this source. The WEBT data are complemented by ultraviolet and X-ray data from {\it Swift}, and by $\gamma$-ray data by {\it Fermi}. 
The big blue bump is modelled by using optical and near-infrared mean spectra obtained during the campaign, together with optical and ultraviolet quasar templates. We give prescriptions to correct the source photometry in the various bands for the thermal contribution, in order to derive the non-thermal jet flux. The role of the intergalactic medium absorption is analysed in both the ultraviolet and X-ray bands. We provide opacity values to deabsorb ultraviolet data, and derive a best-guess value for the hydrogen column density of $N_{\rm H}^{\rm best}=6.3 \times 10^{20} \rm \, cm^{-2}$ through the analysis of X-ray spectra.
We estimate the disc and jet bolometric luminosities, accretion rate, and black hole mass.
Light curves do not show persistent correlations among flux changes at different frequencies. We study the polarimetric behaviour and find no correlation between polarisation degree and flux, even when correcting for the dilution effect of the big blue bump. Similarly, wide rotations of the electric vector polarisation angle do not seem to be connected with the source activity.
\end{abstract}




\section{Introduction}

Blazars are radio-loud active galactic nuclei (AGN) with the peculiarity that one of the relativistic plasma jets points toward us.
The jet emission undergoes Doppler beaming, with consequent flux enhancement, contraction of the variability time scales and blue-shift of the radiation. The blazar spectral energy distribution (SED), in the usual $\log (\nu F\nu)$ versus $\log \nu$ plot, presents two bumps. The low-energy bump is produced by synchrotron radiation, while the origin of the high-energy bump is debated. According to leptonic models, high-energy photons are obtained through inverse-Compton scattering of soft photons on relativistic electrons, while in the hadronic scenario they are produced by acceleration of protons and/or particle cascades.

The blazar class includes BL Lac objects (BL Lacs) and flat-spectrum radio quasars (FSRQs), which were originally distinguished according to the equivalent width of their emission lines \citep{stickel1991}.
Following the unified scheme of AGN, the parent population of blazars are radio galaxies, with FRI and FRII grossly representing the unbeamed counterpart of BL Lacs and FSRQs, respectively \citep{urry1995}. 
Understanding the affinity between blazars and other AGN classes can benefit from the study of the properties of the unbeamed blazar emission coming from the accretion disc and broad line region (BLR). 
This is an extremely difficult task for BL Lacs, which usually have featureless spectra and no signature of disc radiation. In contrast, the spectra of FSRQs generally show broad emission lines. Moreover, their SED displays a big blue bump, and sometimes also a little blue bump, which are interpreted as contributions from the accretion disc and BLR and can reveal important information on the AGN nuclear zone, i.e.\ their quasar core.

One of the most promising candidates to study the unbeamed properties of blazars is the FSRQ 0836+710 (4C~71.07), whose SED shows a particularly luminous disc \citep{raiteri2014}. Its redshift was estimated to be $z=2.172$ by \citet{stickel1993} from the broad emission lines CIV $\lambda1549$ and CIII] $\lambda1909$, and $z=2.18032$ by \citet{lawrence1996}, while \citet{mcintosh1999} derived a systemic redshift of $z=2.218$ from the [OIII] $\lambda5007$ narrow line in $H$-band spectra. 
A detailed investigation of the spectroscopic properties of 4C~71.07 is presented in \citet{raiteri2019}. They estimated a systemic redshift of $z=2.213$ from the Balmer H$\alpha$ and H$\beta$ broad emission lines. In the following, we will adopt this redshift value.

\citet{asada2010} inferred a Faraday rotation measure gradient from multi-frequency VLBI polarimetry, suggesting a helical magnetic field for the jet of 4C~71.07. 
Evidences in favour of a helical jet structure were presented also by \citet{perucho2012_apj} based on very long baseline interferometry data. From the absence of a hot-spot in the arc-second jet radio structure \citet{perucho2012_aa} concluded that the jet likely loses collimation and gets disrupted by the growth of helical instabilities.

\citet{akyuz2013} analysed the multifrequency behaviour of the source during both a quiescent state in 2008--2011 and an active state in 2011. They found that the $\gamma$-ray emission correlates with the optical, but not with the radio emission and that the $\gamma$-ray spectrum becomes curved in active states.

From the theoretical side, in their analys of high-redshift blazars, \citet{ghisellini2010} applied a simple leptonic one-zone synchrotron and inverse-Compton model to the sources SED to derive information on the nuclear and jet physical parameters. For 4C~71.07 they estimated a black hole mass of $3 \times 10^9 \, M_\odot$, a size of the BLR of $1.5 \times 10^{18} \, \rm cm$, an accretion disc luminosity of $2.25 \times 10^{47} \, \rm erg \, s^{-1}$, a bulk Lorentz factor of 14 at the jet dissipation radius of $5.40 \times 10^{17} \, \rm cm$ for a jet viewing angle of 3\degr.

With the aim of disentangling the beamed from the unbeamed properties of this distant FSRQ to study both the jet and nuclear emission, we organised an intense multiwavelength monitoring effort in the period going from September 2014 to October 2016.
Optical (photometric and polarimetric), near-infrared and radio monitoring was obtained by the GLAST-Agile Support Program (GASP) of the Whole Earth Blazar Telescope Collaboration\footnote{http://www.oato.inaf.it/blazars/webt} \citep[e.g.][and references therein]{villata2008,raiteri2017_nature}. 
These observations were complemented by pointings of the {\it Swift} satellite approximately once a month, by optical spectroscopic monitoring at the 4.2-m William Herschel Telescope (WHT) and 2.5-m Nordic Optical Telescope (NOT), and by near-infrared spectroscopic observations at the 3.58-m Telescopio Nazionale Galileo (TNG), all in the Canary Islands, Spain. The continuous survey of the sky at $\gamma$ rays by the {\it Fermi} satellite completed the observing coverage at high energies. 
The $\gamma$-ray flaring activity in 2015 October--November detected by the {\it Astrorivelatore Gamma ad Immagini LEggero} ({\it AGILE}) and by {\it Fermi} has been analysed in \citet{vercellone2019}.
A detailed investigation of the broad emission line properties is presented by \citet{raiteri2019}. 
In this paper, we analyse the photometric and polarimetric data acquired during the WEBT campaign together with the UV and X-ray data from {\it Swift} and $\gamma$-ray data from {\it Fermi}. 

The paper is organised as follows: we first present and analyse in detail the radio-to-ultraviolet data acquired from both ground-based and space observations (Sections from 2 to 5). These data are used in Section 6 to reconstruct the low-energy bump of the source SED, to disentangle the synchrotron from the nuclear thermal emission, and to build an empirical model for the latter. We subsequently analyse the high-energy X-ray (Section 7) and $\gamma$-ray (Section 8) data and discuss the broad-band multiwavelength behaviour (Section 9). We finally present optical polarimetric observations (Section 10) and draw conclusions in Section 11.

\section{Ground-based optical, near-infrared and radio data}
\label{webt}

Optical observations for the WEBT campaign were performed at the following observatories:
Abastumani (Georgia), 
Belogradchik (Bulgaria), 
Calar Alto\footnote{Calar Alto data were acquired as part of the MAPCAT project: http://www.iaa.es/$\sim$iagudo/research/MAPCAT} (Spain),
Crimean (Russia), 
Lowell (USA; Perkins telescope), 
Lulin (Taiwan),
McDonald\footnote{In the framework of the telescope network of the Las Cumbres Observatory.}  (USA),
Mt.\ Maidanak (Uzbekistan), 
Pulkovo (Russia), 
Roque de los Muchachos (Spain; Liverpool, NOT, TNG, and WHT telescopes), 
ROVOR (USA), 
Rozhen (Bulgaria), 
SAI Crimean (Russia), 
Sirio (Italy),
Skinakas (Greece), 
St.\ Petersburg (Russia), 
Teide (Spain; IAC80 and STELLA-I telescopes),
Tijarafe (Spain), 
Valle d'Aosta (Italy),
Vidojevica Astronomical Station (Serbia; 60 and 140 cm telescopes).

\begin{figure}
	\includegraphics[width=\columnwidth]{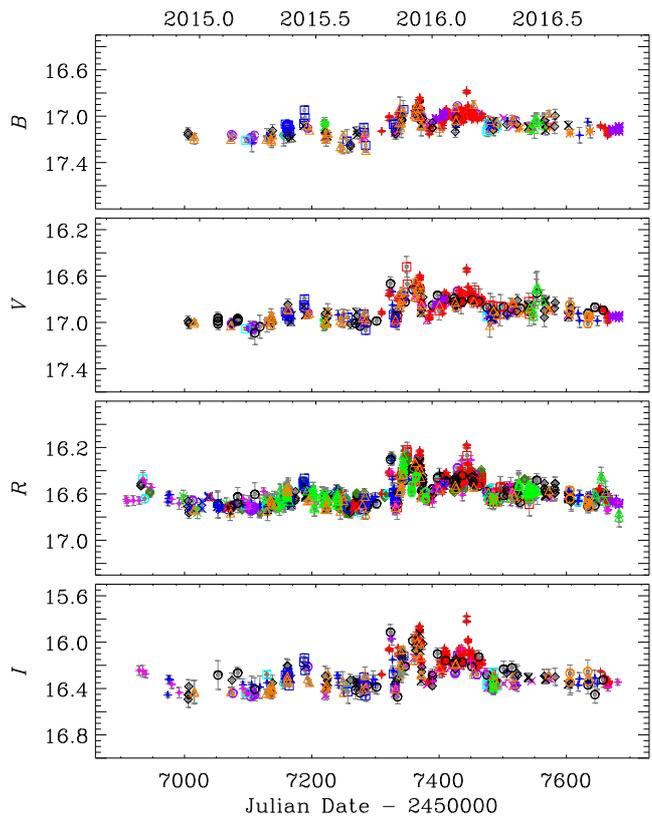}
    \caption{Optical light curves of 4C~71.07 from WEBT observations: observed magnitudes versus Julian Date ($-2450000$). Different datasets are marked with different colours and symbols.}
    \label{webt_opt}
\end{figure}

\begin{figure}
	\includegraphics[width=\columnwidth]{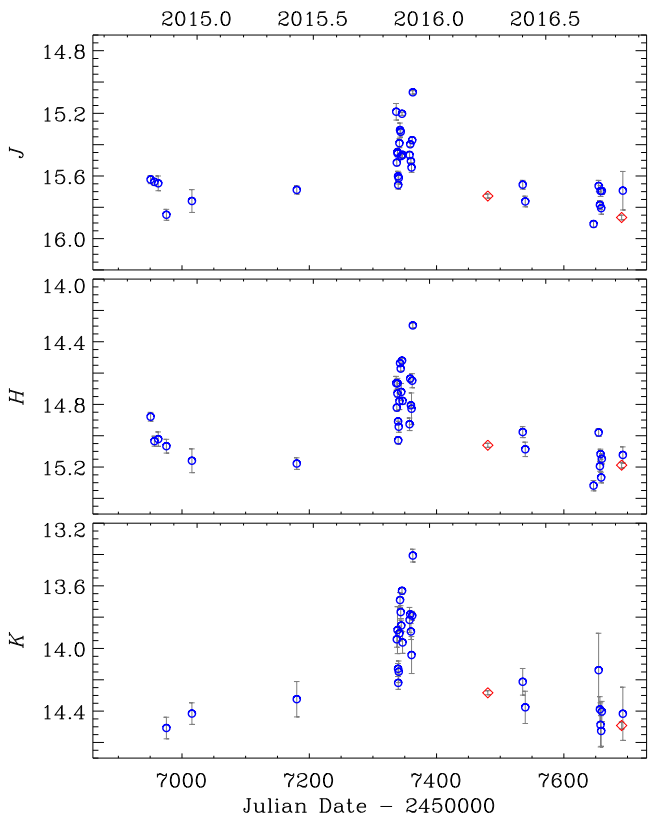}
    \caption{Near-infrared light curves of 4C~71.07 from WEBT observations: observed magnitudes versus Julian Date ($-2450000$). Blue circles represent data from Campo Imperatore, red diamonds observations taken with the TNG at the Roque de los Muchachos Observatory.}
    \label{webt_nir}
\end{figure}

All datasets were processes with standard procedures. The source magnitude was obtained by differential aperture photometry, using reference stars in the same field of the source \citep[][Larionov, private communication]{villata1997,doroshenko2014}.
Further optical photometry was obtained as calibration information in support of the spectroscopic monitoring at the NOT and WHT telescopes \citep[see][]{raiteri2019}.

Overposition of data from different telescopes sometimes revealed offsets that were corrected for by shifting the most deviating datasets. Strong outliers showing up in only one band were removed. In some cases data scatter was reduced through binning.

Near-infrared data in $J$, $H$, and $K$ bands were acquired at the Campo Imperatore Observatory (Italy). Additional data were obtained as calibration photometry for the near-infrared TNG spectra \citep[see][]{raiteri2019}.

The final, cleaned, optical and near-infrared light curves are shown in Figs.\ \ref{webt_opt} and \ref{webt_nir}, respectively.
Notwithstanding the large sampling difference, especially between optical and near-infrared data, and different precision, the general increase of the variability amplitude from blue to red is evident and reveals the imprint of the emission contribution from the big blue bump, which is stronger in the blue (see Section \ref{template}). Table \ref{dati} reports, for each band, the number of data in the final light curves, the variability amplitude $\Delta \rm mag=mag_{\rm max}-mag_{\rm min}$ and the mean fractional variation \citep{peterson2001}
$$f_{\rm var}={{\sqrt{\sigma^2-\delta^2}} \over {<F>}},$$
where $\sigma^2$ is the dataset variance, $\delta^2$ the mean square uncertainty of the fluxes (see Section \ref{flux} for the transformation from magnitudes to fluxes), and $<F>$ is the mean flux of the dataset.
The advantage of $f_{\rm var}$ is that it takes into account the data errors.
A few flares are visible in the optical light curves in the period from $\rm JD=2457320$ to  $\rm JD=2457370$ 
and then at $\rm JD=2457443$. The optical maxima were missed by the near-infrared observations, so we can expect that the near-infrared variability amplitude were actually higher than reported in Table \ref{dati}.

\begin{table}
	\centering
	\caption{Source properties in the various photometric bands: number of data, variability amplitude, mean fractional variation, big blue bump contribution to the flux densities corrected for the Galactic extinction.}
	\label{dati}
	\begin{tabular}{cccccc}
\hline
Band & $N_{\rm data}$ & $\Delta \rm mag$ & $f_{\rm var}$ & $F_{\rm BBB}$ (mJy) & $\log (\nu F_\nu)_{\rm BBB}$\\
\hline
\multicolumn{6}{c}{WEBT}\\
\hline
$B$  &  470 & 0.617 & 0.07 & 0.527 & -11.443\\
$V$  &  531 & 0.561 & 0.09 & 0.520 & -11.543\\
$R$  & 1521 & 0.717 & 0.10 & 0.532 & -11.604\\
$I$  &  507 & 0.715 & 0.13 & 0.506 & -11.721\\
$J$  &   36 & 0.841 & 0.19 & 0.470 & -11.943\\
$H$  &   36 & 1.023 & 0.23 & 0.418 & -12.123\\
$K$  &   30 & 1.120 & 0.28 & 0.606 & -11.075\\
\hline
\multicolumn{6}{c}{UVOT}\\
\hline
$w2$ & 43 & 0.47 & 0.10 & 0.045 & -12.200 \\
$m2$ & 42 & 0.48 & 0.08 & 0.059 & -12.108 \\
$w1$ & 43 & 0.48 & 0.10 & 0.137 & -11.815 \\
$u$  & 43 & 0.38 & 0.07 & 0.421 & -11.442 \\
$b$  & 43 & 0.35 & 0.07 & 0.533 & -11.438 \\
$v$  & 42 & 0.55 & 0.06 & 0.513 & -11.548 \\
\hline
\end{tabular}
\end{table}

In the radio band, observations were performed at the Pico Veleta\footnote{These data were acquired with the IRAM 30 m telescope as part of the POLAMI (Polarimetric AGN Monitoring with the IRAM-30 m-Telescope) and MAPI (Monitoring AGN with Polarimetry at the IRAM 30 m Telescope) programmes.} (Spain; 228 and 86 GHz),  Mets\"ahovi (Finland; 37 GHz), and Medicina (Italy; 24, 8, and 5 GHz) observatories. Data reduction and analysis procedures are described in \citet{agudo2010}, \citet{terasranta1998} and \citet{dammando2012}. Radio light curves will be shown in Fig.\ \ref{mw}. The 37 GHz data show some scatter, so we plotted a cubic spline interpolation through the 30-binned data to better distinguish the trend.

\section{Ultraviolet and optical observations by {\it Swift}-UVOT}

In the period of the WEBT campaign, {\it Swift} pointed at the source during 43 epochs.
We processed the data with Heasoft version 6.22.
The source counts were extracted from a circular region with 5 arcsec radius centred on the source; the background counts were derived from an annular region centred on the source with inner and outer radius of 10 and 20 arcsec, respectively.

\begin{figure}
	\includegraphics[width=\columnwidth]{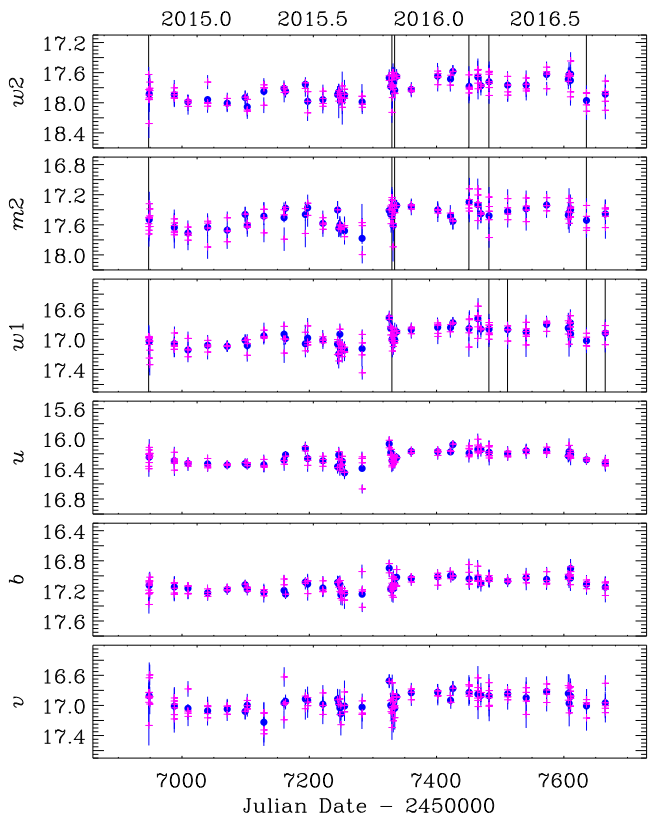}
    \caption{Ultraviolet and optical light curves built with data from the UVOT instrument onboard {\it Swift}. Magenta plus signs represent photometry derived from the single exposures; blue dots that obtained by summing the frames in the same filter at the same epoch. Vertical lines mark the epochs affected by small scale sensitivity problems.
}
    \label{uvot}
\end{figure}

We processed both the single exposures and the images obtained by summing the exposures with the same filter in the same epoch. The results are shown in Fig.\ \ref{uvot}.

Observations were checked for small scale sensitivity (sss) problems, which occur when the source falls on small detector regions where the sensitivity is lower. The problem is more important for the ultraviolet filters.
We found 35 snapshots in 8 epochs where sss effects are recognised by the check procedure\footnote{http://swift.gsfc.nasa.gov/analysis/uvot\textunderscore digest/sss\textunderscore check.html}, all in the ultraviolet filters.
These epochs are shown in Fig.\ \ref{uvot}. The data dispersion there is not larger than what happens in other epochs, so there is no need to exclude some of these frames from the analysis. Indeed, the most questionable points, lying out of the general trend common to all filters (e.g. the $v$-band point at $\rm JD=2457128$) are not due to sss problems. 

As shown by Fig.\ \ref{uvot} and Table \ref{dati}, the source variability is smaller in the UVOT optical bands than in the ultraviolet, and this is a consequence of the fact that, due to the high redshift of the source, the big blue bump peaks in the $u$--$b$ bands (see Section \ref{template}).

The comparison with the ground-based data shown in Fig.\ \ref{webt_opt} reveals that only the optical flaring period around $\rm JD=2457330$ is well covered by UVOT observations, which were triggered by the detection of a high $\gamma$-ray flux \citep{vercellone2019}, while the other optical peaks were missed. As a consequence, the variation amplitude and mean fractional variability reported in Table \ref{dati} for the UVOT filters underestimate the actual variability of the source in the considered period.

\section{Colour behaviour}

To investigate the source spectral behaviour, we first built ground-based $B-V$ colour indices by associating the most precise $B$ and $V$ data (error less than 0.03 mag) acquired within 15 minutes by the same telescope. We obtained 347 values, with an average index of 0.19 and standard deviation of 0.03. 

\begin{figure}
	\includegraphics[width=\columnwidth]{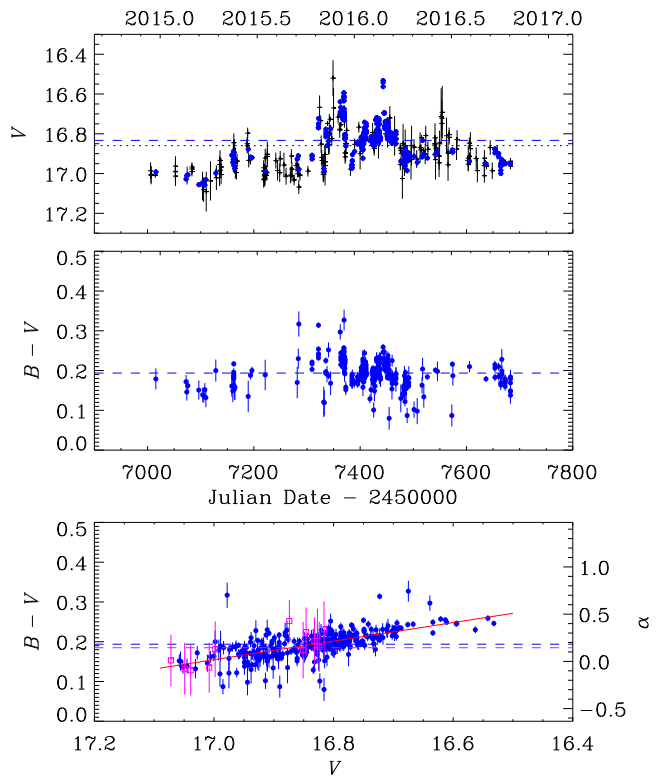}
    \caption{$B-V$ colour indices as a function of time (middle panel) and brightness (bottom panel); the dashed blue line indicates the average value. The light curve in $V$ band is shown in the top panel for comparison. Black plus signs and dotted line refer to the whole dataset; blue dots and dashed line to the subset used to get colour indices.The small difference in $<V>$ between the whole dataset and the subset means that colour indices sample fairly well the range of brightness covered by the source. In the bottom panel, the linear regression line is plotted in red. The spectral index $\alpha$ of the $F_\nu \propto \nu^{-\alpha}$ law is shown on the right. Magenta squares and lines refer to UVOT data.}
    \label{colori}
\end{figure}

Figure \ref{colori} shows that the colour indices sample the whole brightness range of the source and clearly indicate a redder-when-brighter behaviour. Linear regression results in a slope of $-0.23$.
This trend is expected if the source brightening is due to the increasing contribution of a ``red" variable synchrotron component to the (quasi) stationary emission of the big blue bump \citep[e.g.][]{gu2006,villata2006}.

In the same figure we show the $b-v$ colour indices obtained from UVOT data. We had to relax the constraints on the errors to get a reasonable number of colours. Using data with uncertainties less than 0.06 mag, we obtained 12 indices. They cover only the faintest states and indicate a mean value of 0.18, with standard deviation of 0.04, in agreement with the ground-based data.

\section{From magnitudes or count rates to fluxes}
\label{flux}

Optical and near-IR magnitudes were transformed into flux densities by correcting for Galactic reddening according to the NASA/IPAC Extragalactic Database\footnote{http://ned.ipac.caltech.edu} (NED) and using the absolute fluxes by \citet{bessell1998}.

By assuming a power-law shape of the optical spectrum $F_\nu \propto \nu^{-\alpha}$, we can derive the optical spectral index from the colours (see Fig.\ \ref{colori}). The mean value obtained from the ground-based $B-V$ indices is $\alpha=0.18$ with a standard deviation of 0.14.

As seen in the previous section, the average $B-V$ or $b-v$ colour is $\sim 0.2$ mag, which is outside the range of validity of the \citet{breeveld2011} count rate to flux conversion factors for the UVOT ultraviolet bands. This means that the standard UVOT calibrations in the ultraviolet are not applicable to spectral types like that of our source. Therefore, to convert UVOT magnitudes into fluxes, we followed the recalibration procedure described in \citet{raiteri2010} and applied to a number of cases thereafter \citep[e.g.][]{dammando2012,larionov2016}.
We convolve an average source spectrum with the UVOT filter effective areas to derive the effective wavelengths $\lambda_{\rm eff}$ and count-to-flux conversion factors $CF$, and further with the mean extinction laws by \citet{cardelli1989} to obtain the extinction values $A_\lambda$. The procedure is then iterated to check stability of the results.
Those reported in Table \ref{recal} were obtained with a log-linear fit to the source spectrum.

\begin{table}
	\centering
	\caption{Results of the UVOT recalibration procedure and prescriptions to correct for both Galactic and IGM absorption.}
	\label{recal}
	\begin{tabular}{cccccc}
\hline
     & $\lambda_{\rm eff}$ & $CF$                                          & $A_\lambda$ & $\tau_{\rm eff}^a$& $\tau_{\rm eff}^b$\\
     & (\AA)               & ($\rm erg \, cm^{-2} \, s^{-1}\, \AA\ ^{-1}$) & (mag)       &                 & \\
\hline
$w$2 & 2148 & $6.009 \, 10^{-16}$ & 0.220 & 1.140 & 1.481 \\
$m$2 & 2272 & $8.358 \, 10^{-16}$ & 0.231 & 0.923 & 1.330 \\
$w$1 & 2688 & $4.405 \, 10^{-16}$ & 0.183 & 0.529 & 0.820 \\
$u$  & 3491 & $1.647 \, 10^{-16}$ & 0.133 & 0.147 & 0.205 \\
$b$  & 4377 & $1.467 \, 10^{-16}$ & 0.111 & 0.    & 0. \\
$v$  & 5439 & $2.602 \, 10^{-16}$ & 0.084 & 0.    & 0. \\
\hline
\multicolumn{6}{l}{$^a$ Derived following \citet{ghisellini2010}}\\
\multicolumn{6}{l}{$^b$ Derived following \citet{lusso2015}}\\
\end{tabular}
\end{table}

With respect to the \citet{breeveld2011} calibrations, the most noticeable differences are a shift of about 100 \AA\ in the effective wavelengths of the $w$1 and $w$2 bands and a $\sim 5\%$ increase of the count-to-flux conversion factor in the $w$1 band.

Starting from the source count rates, we applied the count-to-flux conversion factors of Table \ref{recal} and then corrected for the Galactic extinction according to the values in the same table to get deabsorbed flux densities. 
However, because of the high redshift of the source, the spectral region blueward of the Ly$\alpha$ emission line is strongly eroded by a wealth of intervening absorbers
\citep{scott2000,raiteri2019}.
Therefore, to reconstruct the flux observed in the ultraviolet as it was emitted from the source, we must further correct for this effect. This issue will be addressed in the next section.

\section{An empirical model for the quasar core}
\label{template}

\begin{figure*}
	\includegraphics[width=15cm]{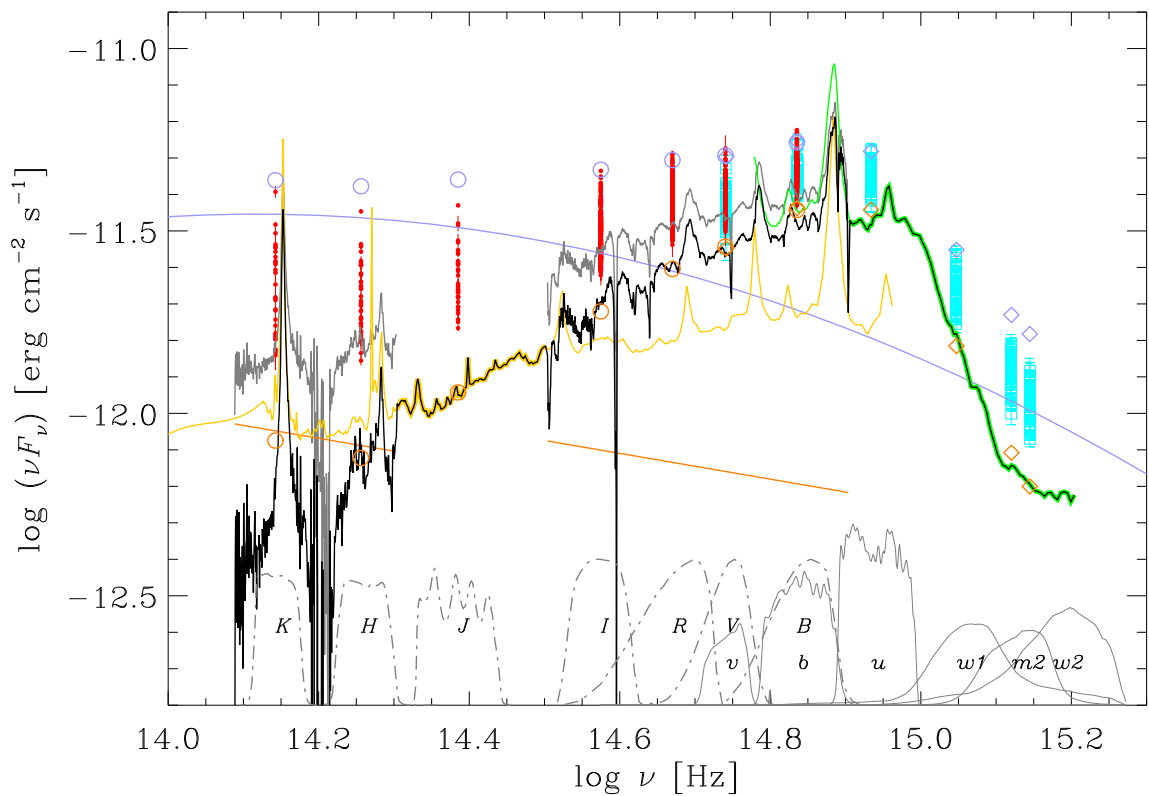}
    \caption{Spectral energy distribution of 4C~71.07 from the near-infrared to the ultraviolet. 
Red dots and cyan squares represent the photometric data acquired by the WEBT and by {\it Swift}-UVOT, respectively. The average near-infrared and optical spectra by \citet{raiteri2019} are shown in grey; the power laws used to correct them for the synchrotron jet emission contribution are plotted as orange lines. The quasar templates by \citet{polletta2007} and \citet{lusso2015} are shown in light orange and green, respectively.
The final empirical model for the 4C~71.07 quasar core is plotted as a black line. Its contributions in the various photometric bands (whose transmission curves are shown in the figure bottom) are marked with orange circles (diamonds for the UVOT bands).
The violet log-parabola represents a flaring state jet emission. The addition of this contribution to the nuclear thermal emission produces the photometric values shown as violet circles (diamonds for the UVOT filters). 
}
    \label{sed_template}
\end{figure*}

\begin{figure}
	\includegraphics[width=\columnwidth]{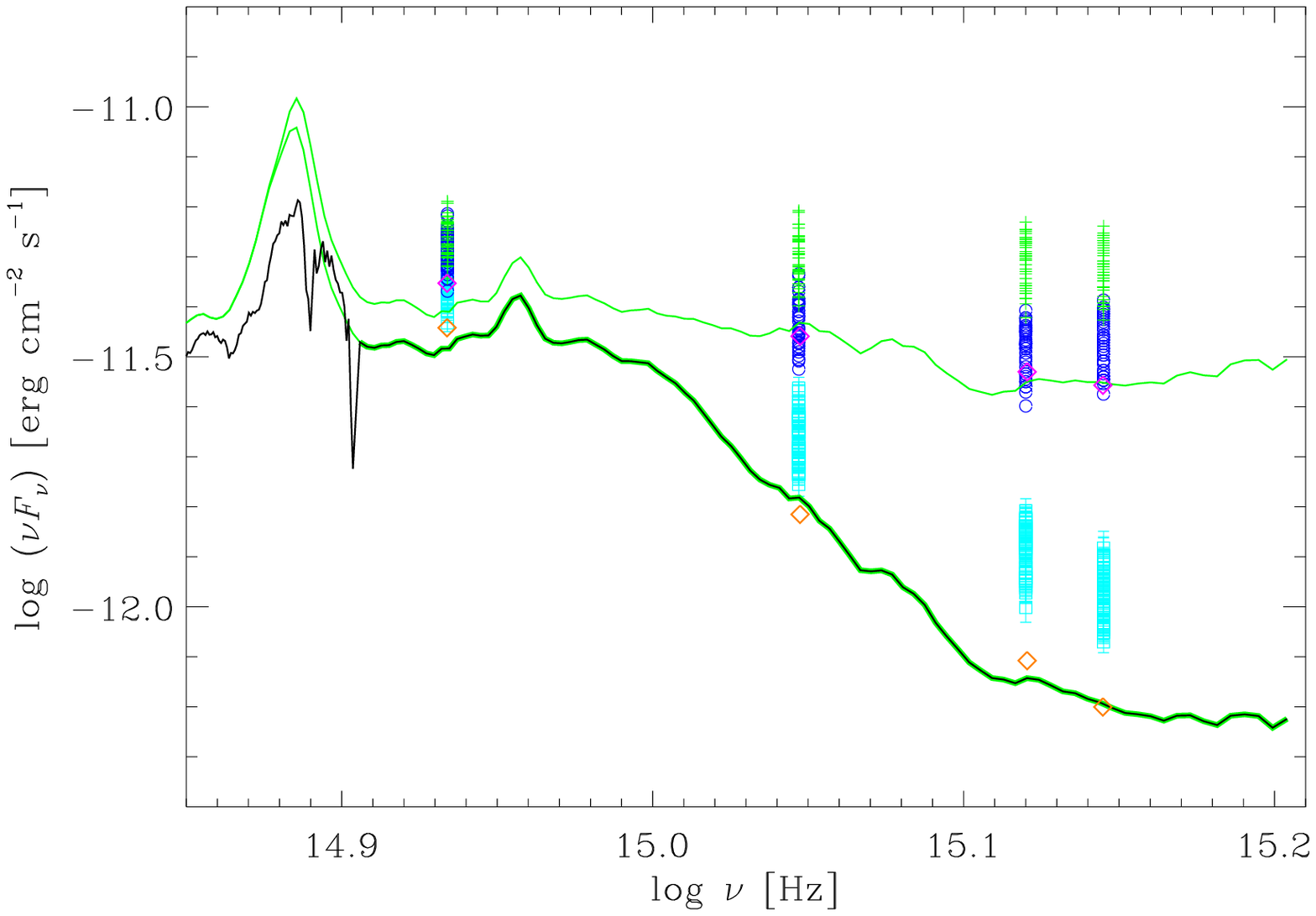}
    \caption{Zoom into the spectral energy distribution of 4C~71.07 in the blue--ultraviolet.
The black line represents the quasar core template, which corresponds to the stacked quasar spectrum of \citet{lusso2015} (lower green line) at wavelengths shorter than 3740 \AA. The upper green line shows the IGM-corrected spectrum by the same authors. Orange diamonds mark the big blue bump contributions in the photometric UVOT bands derived from the absorbed template. The magenta diamonds show the big blue bump contributions after correction for the IGM absorption using the IGM-corrected template. Cyan squares represent the UVOT data, while blue circles and green plus signs the same data corrected for IGM absorption with the \citet{ghisellini2010} and \citet{lusso2015} prescriptions, respectively.
}
    \label{sed_igm}
\end{figure}

Based on infrared data from the {\it Wide-field Infrared Survey Explorer} ({\it WISE}) satellite, the Two Micron All-Sky Survey (2MASS), and the Campo Imperatore and Teide observatories, \citet{raiteri2014} modelled the SED of 4C~71.07 from the infrared to the ultraviolet with the superposition of a log-parabolic jet component and a nuclear thermal component representing the emission contribution of the accretion disc and broad line region. Because of the prominence of the big blue bump in this source, the nuclear thermal model was obtained by strengthening the type 1 QSO template by \citet{polletta2007} with a black body.
Thanks to the wealth of photometric and spectroscopic data we obtained in the WEBT two-year campaign, we can now refine the model for the source quasar core and estimate the thermal contribution of the nuclear emission to the source photometry in the various bands.

Fig.\ \ref{sed_template} shows the near-infared to ultraviolet SED of 4C 71.07. All photometric and spectroscopic data have been corrected for the Galactic absorption.
To build the core template, we started by considering the range of brightness spanned by the near-infrared, optical, and ultraviolet photometric data in the period considered in this paper. As mentioned before, the variability amplitude increases towards the red, and this is due to the increasing contribution of the very variable synchrotron emission with respect to the less variable thermal emission, which is assumed to be steady in the relatively short period of time we are dealing with. 

In \citet{raiteri2019} we present and discuss in detail the results of the spectroscopic monitoring of 4C~71.07 during the WEBT campaign.
We obtained 24 optical spectra (12 with the WHT and 12 with the NOT), and 2 near-infrared spectra with the TNG. All of them were carefully calibrated in flux also using photometric supporting data.
The average optical and near-infrared spectra are shown in Fig.\ \ref{sed_template}. They have not been corrected for atmospheric absorption and include both a non-thermal emission contribution from the jet and a thermal contribution from the quasar core. To obtain a template for the big blue bump, we must first clean the spectra from the jet component, which we model as a power-law. We use the relative difference in flux variability in the various bands to determine the slope of the power-law and set the brightness level 
to have a very strong thermal contribution to the $B$ and $V$-band fluxes, as suggested by their observed smaller variability (see Figs.\ \ref{webt_opt}--\ref{uvot} and Table \ref{dati}).
Although the model normalisation is somewhat arbitrary, the log-parabolic shape of the thermal-subtracted SED that we will discuss in Section 9 suggests that we are not far from the real, elusive solution.
By comparing the spectroscopic with the photometric information, we note that the source brightness level corresponding to the near-infrared spectrum is lower than that corresponding to the optical spectrum, therefore we use a power-law with a lower normalisation to describe the jet contribution to the near-infrared spectrum.
By subtracting the jet flux from the spectral fluxes, we obtain the predicted nuclear component.

We lack source spectra in between the near-infrared and the optical spectrum as well as in the ultraviolet. Therefore, we complete the nuclear thermal model using available quasar templates.
We adopted the TQSO1 template by \citet{polletta2007}, shifted to the systemic redshift of $z=2.213$ \citep{raiteri2019} and properly rescaled to smoothly join our optical and near-infrared spectra, to cover the wavelength range from 14900 to 9400 \AA. 
The comparison between the prolongation of the TQSO1 template and the optical spectrum of 4C~71.07 corrected for the jet emission (see Fig.\ \ref{sed_template}) reveals that the rising part of the big blue bump of this source is much harder that the average quasar spectrum, as found by \citet{raiteri2014}. This may be due to the fact that in other AGN the low-frequency part of the disc SED is usually contaminated by additional softer emission contributions, as suggested by \citet{calderone2013}.

For wavelengths shorter than 3740 \AA\ we use the ultraviolet quasar-stacked spectrum by \citet{lusso2015}. This was obtained by combining spectra of 53 quasars with redshift $z \sim 2.4$ acquired with the WFC3 instrument of the {\it Hubble Space Telescope} (HST).
The authors present both the observed spectrum and that obtained by correcting for the absorption by the intergalactic medium (IGM). We use the former (Lusso, private communication), properly scaled, to complete our ``observed" big blue bump template, while the IGM-corrected spectrum will be used to estimate effective opacity values in the photometric bands blueward of the Ly$\alpha$.

The final empirical model for the quasar core emission of 4C~71.07 is shown in Fig.\ \ref{sed_template}.
By convolving this template with the transmission curve of the Bessels and UVOT filters, we then calculate the photometric contributions of the big blue bump in the various bands\footnote{We had to apply a correction to the $w2$ values to take into account that the template does not cover the whole range of frequencies spanned by the $w2$ filter. The correction was done by shifting the $w2$ thermal contribution to match the template.}. 
These are reported in Table \ref{dati} and can be used to subtract the thermal contributions to the source photometric observations when the purely non-thermal, jet emission is desired.

To further check the consistency of our procedure, we simulated a high brightness state. The broad-band jet emission is now modelled with a log-parabola, which is often used to describe the synchrotron contribution \citep[e.g.][]{massaro2004,raiteri2017_nature}, as the power-law approximation would be too rough over such a large frequency range. By summing the log-parabola and nuclear thermal template fluxes, we obtain what we should observe in flaring states. These predictions can then be compared to the highest observed flux levels. 
In Fig.\ \ref{sed_template} we see that the optical maxima are satisfactorily reproduced, while the near-infrared maxima are somewhat overproduced. This can be at least partly justified by the fact that we lack near-infrared data at the epochs of the optical flux peaks (see Section \ref{webt}).
In the ultraviolet, the range of the observed data is satisfactorily reproduced too. 
These are encouraging results, especially when considering that the ultraviolet template we used \citep{lusso2015} is an average quasar spectrum and that the jet emission spectrum can likely only approximately be described with a log-parabolic model. 

A further step is now necessary to correct for the IGM absorption.
Effective opacity values due to such absorption were estimated by \citet{ghisellini2010} by averaging over all possible lines of sight. We rescaled those values (Ghisellini, private communication) to take into account the difference between the old standard effective wavelengths by \citet{poole2008} and ours. These new effective opacity values $\tau_{\rm eff}$ are listed in Table \ref{recal}. With respect to the estimates of \citet{ghisellini2010} there is a 6\% decrease in the $w2$ band and a 2\% decrease in the $m2$ and $w1$ bands. The IGM-corrected flux densities are obtained as $F_\tau=F \times \exp(\tau_{\rm eff})$ and are shown in Fig.\ \ref{sed_igm}. 

Another way of estimating the correction is to use the IGM-corrected quasar spectrum of \citet{lusso2015}.
By convolving this spectrum with the UVOT filters effective areas, we obtain the photometric contributions of the deabsorbed quasar core of 4C~71.07 (see Fig.\ \ref{sed_igm}). 
The differences between deabsorbed and absorbed SED values\footnote{The ratio $T_{\lambda}=F_{\lambda, \rm obs}/F_{\lambda, \rm corr}$ corresponds to the mean IGM transmission function of \citet{lusso2015}.} allow us to estimate the average effective opacities and to correct the UVOT data for IGM absorption. The opacity values are reported in Table \ref{recal}, while corrected data are shown in Fig.\ \ref{sed_igm}. As can be seen, following \citet{lusso2015} leads to a higher correction than estimated by \citet{ghisellini2010}.

A final comment is due to remind that the disc emission of quasars is variable and also in the case of 4C~71.07 a remarkable change (a flux variation of a factor $\sim 2.5$) has been noted by \citet{raiteri2019}. However, this occurred on a time span of more than 30 years and indeed quasar flux changes are usually of the order of a few tenths of mag on time scales of several months/years \citep[e.g.][]{kaspi2000}. This is much less than the variability characterising the non-thermal radiation from the jet.

\section{Disc luminosity, accretion rate and black hole mass}
\begin{figure}
	\includegraphics[width=\columnwidth]{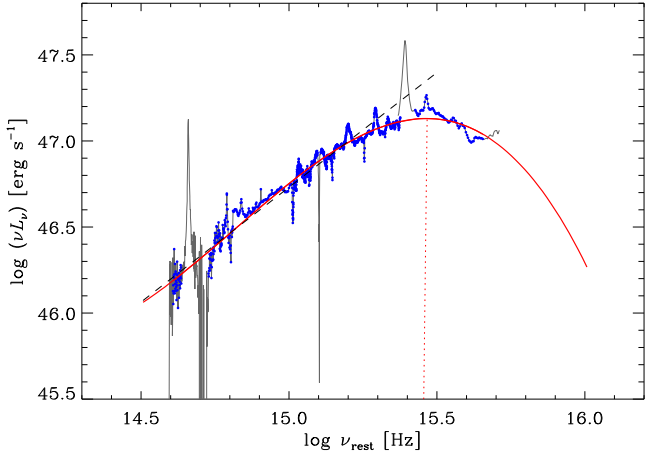}
    \caption{The empirical template for the 4C~71.07 big blue bump corrected for IGM absorption (grey). Blue dots mark the points that were used to obtain a third-order polynomial fit to the thermal continuum (red thick line). The dotted vertical line highlights the peak. 
The black dashed line represents the $F_{\nu} \propto \nu^{1/3}$ slope of a Shakura \& Sunyaev disc spectrum.}
    \label{lbol}
\end{figure}

The empirical model built in the previous section allows us to estimate the disc bolometric luminosity by integrating the thermal continuum. This is obtained by fitting the big blue bump template with a cubic polynomial (see Fig.\ \ref{lbol}) and leads to the extremely high value $L_{\rm disc}=2.45 \times 10^{47} \rm \, erg \, s^{-1}$, assuming a luminosity distance of 17585 Mpc. The main uncertainty comes from the high-energy part of the spectrum, which is poorly constrained. However, our estimate is in good agreement with the value $2.25 \times 10^{47} \rm \, erg \, s^{-1}$ calculated by \cite{ghisellini2010} with a completely different procedure.

The peak of the thermal continuum is found at $\log \nu_{\rm rest} \simeq 15.46$ and implies a peak luminosity of $(\nu L_\nu)_{\rm peak}=1.35 \times 10^{47} \rm \, erg \, s^{-1}$, so that $L_{\rm disc}/(\nu L_\nu)_{\rm peak}=1.8$, close to the factor 2 usually assumed \citep[e.g.][]{ghisellini2015,calderone2013}. Moreover, the peak frequency is in agreement with that predicted by accretion disc models for the same luminosity \citep[e.g.][]{hubeny2000}.

We note that the rising part of our disc template fairly matches the $F_{\nu} \propto \nu^{1/3}$ spectral distribution of a Shakura \& Sunyaev disc \citep{shakura1973} up to $\log \nu_{\rm rest} \sim 15.25$.

From the equation $L_{\rm disc}=\eta \dot{M} c^2$ \citep{shakura1973} we can derive the accretion rate $\dot{M}$ once the efficiency of gravitational energy release  $\eta$ is fixed.  This can be as small as 0.06 for a Schwarzschild's black hole, and up to 0.32 for a rotating Kerr's black hole \citep{calderone2013}. The disc bolometric luminosity we estimated above then leads to accretion rates of 17 and 3.3 $\rm M_\odot \, yr^{-1}$, respectively.

\citet{calderone2013} proposed a method to estimate the black hole mass and accretion rate from the disc luminosity based on a Shakura \& Sunyaev disc. By applying their equations (8) and considering that the isotropic disc luminosity is about one half the bolometric luminosity, we obtain $M_{\rm BH}=(1.61$--$1.62) \times 10^9 \, \rm M_\odot$ and $\dot{M}=(18.0$--$18.3) \rm \, M_\odot \, yr^{-1}$ for viewing angles in the range 0\degr--10\degr, as expected for a blazar. 
The Eddington luminosity is then $L_E=1.6 \times 10^{47} \rm \, erg \, s^{-1}$ and the Eddington ratio is 0.66, a remarkably high value \citep[e.g.][]{ghisellini2015}.
The black hole mass is somewhat smaller than that derived by \citet{ghisellini2010}, who found $3 \times 10^9 \, \rm M_\odot$, but the difference is less than a factor 2, which is the expected uncertainty on the results.

A comparison of the nuclear properties estimated above with those inferred from the analysis of the broad emission lines is deferred to \citet{raiteri2019}.

\section{{\it Swift}-XRT}

XRT data were processed with version 6.24 of the {\tt HEAsoft}\footnote{https://heasarc.nasa.gov/lheasoft/} package and calibration files dated 20180710. We ran the {\tt xrtpipeline} on all observations in pointing mode in the period of interest and ended up with 21 observations in WT mode and 43 observations in PC mode. Because all WT observations have less than 1 min exposure, in the following we concentrate on the observations in PC mode. Many of them are piled-up, and the analysis of the source point spread function with the {\tt ximage} tool indicates that the problem affects the inner 3 pixel radius core (1 pixel=2.36 arcsec). To correct for pile-up it is necessary to puncture the centre of the region from which the source counts are extracted and to reconstruct the PSF central maximum from the wings. To this aim, we first run the {\tt xrtcentroid} tool to accurately identify the source coordinates on each image. Then, we extracted the source counts from an annulus with 3 and 30 pixel radii and the background counts in an annulus with 40 and 60 pixel radii centred on the source.

The source spectra, grouped in at least 20 counts per energy bin, were analysed in the 0.3--10 keV energy range with the Xspec package. We adopted the \citet{wil00} elemental abundances and a value for the Galactic absorption of $N_{\rm H}=2.76 \times 10^{20} \rm \, cm^{-2}$ \citep{kal05}.

The XRT data examined by \citet{ghisellini2010} were modelled with a power law with Galactic absorption and the fit was very good from a statistical point of view ($\chi^2/\nu=0.99$). Similarly, an analysis of the X-ray data acquired by {\it XMM-Newton} in 2001 by \citet{vercellone2019} found only marginal evidence for absorption larger than the Galactic value; when left free, the hydrogen column density resulted in $N_{\rm H}=(3.3 \pm 0.2) \times 10^{20} \rm \, cm^{-2}$. They fit the XRT data taken in 2015 with an absorbed power law with $N_{\rm H}$ both free and fixed to the Galactic value. In the former case, values between 3.1 and $7.4 \times 10^{20} \rm \, cm^{-2}$ were obtained as well as higher spectral indices, i.e.\ softer spectra. \citet{arcodia2018} studied the IGM absorption towards high-redshift blazars. They applied different models to the X-ray spectra of several sources, and concluded that the best results are obtained when assuming that the intrinsic spectrum is curved (e.g.\ a log-parabola) and there is some extra-absorption. They also noted that 4C~71.07 is an outlier in the $N_{\rm H}(z)$ versus $z$ relation, showing smaller excess absorption than expected.

\begin{figure}
	\includegraphics[width=\columnwidth]{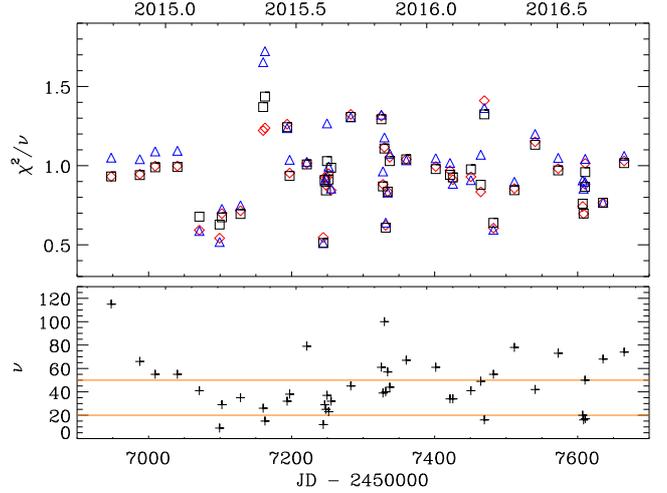}
    \caption{Top: Reduced chi-squared values obtained by fitting the XRT spectra with: a power law with Galactic absorption (blue triangles), a power law with free absorption (red diamonds), and a power law with absorption fixed to our best-guess value $N_{\rm H}^{\rm best}=6.3 \times 10^{20} \rm \, cm^{-2}$ (black squares). Bottom: number of degrees of freedom in the fixed absorption cases.}
    \label{xrt_chi2}
\end{figure}

\begin{figure}
	\includegraphics[width=\columnwidth]{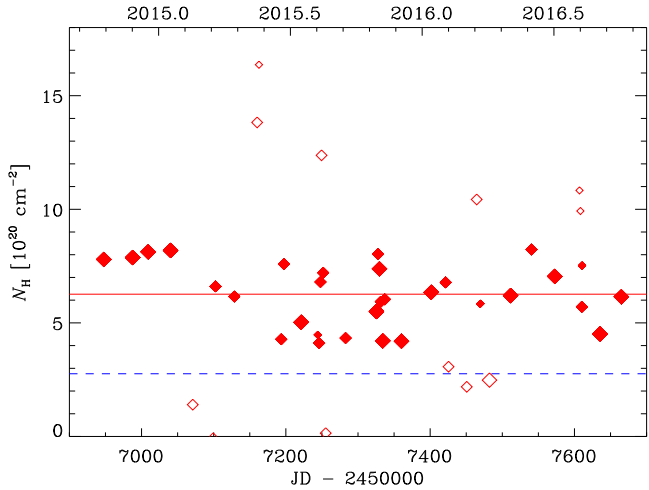}
    \caption{Values of the hydrogen column density obtained by fitting the XRT spectra with a power law with free $N_{\rm H}$. The red horizontal line indicates the best-guess value $N_{\rm H}^{\rm best}=6.3 \times 10^{20} \rm \, cm^{-2}$, while the blue dashed line marks the Galactic value. Symbols are shown with increasing size for increasing degrees of freedom within the ranges indicated by the orange horizontal lines in Fig.\ \ref{xrt_chi2}. Filled symbols highlight those values that are 1 standard deviation within the mean and that have been considered to estimate the best-guess $N_{\rm H}$.}
    \label{xrt_nh}
\end{figure}

In the previous section we saw that the UV emission of 4C~71.07 is likely to be strongly absorbed by the IGM. Therefore, we need to carefully investigate the role of absorption also at X-ray energies.
We first compared the results obtained by fitting the XRT spectra with a power law with Galactic absorption to those obtained when $N_{\rm H}$ is left free to vary. Fig.\ \ref{xrt_chi2} shows that the improvement of the goodness-of-fit in the latter case is substantial, especially for some of the spectral fits. The best-fit $N_{\rm H}$ values are plotted in Fig.\ \ref{xrt_nh}. They show a large range of values, with a mean $N_{\rm H}=6.4 \times 10^{20} \rm \, cm^{-2}$ and a standard deviation of $\sigma=3.3 \times 10^{20} \rm \, cm^{-2}$. If we discard the cases that are $1 \, \sigma$ out from the mean, we obtain $N_{\rm H}^{\rm best}=6.3 \times 10^{20} \rm \, cm^{-2}$. We consider this value as the best guess we can make for the total absorption affecting the X-ray spectra of 4C~71.07.

\begin{figure}
	\includegraphics[width=\columnwidth]{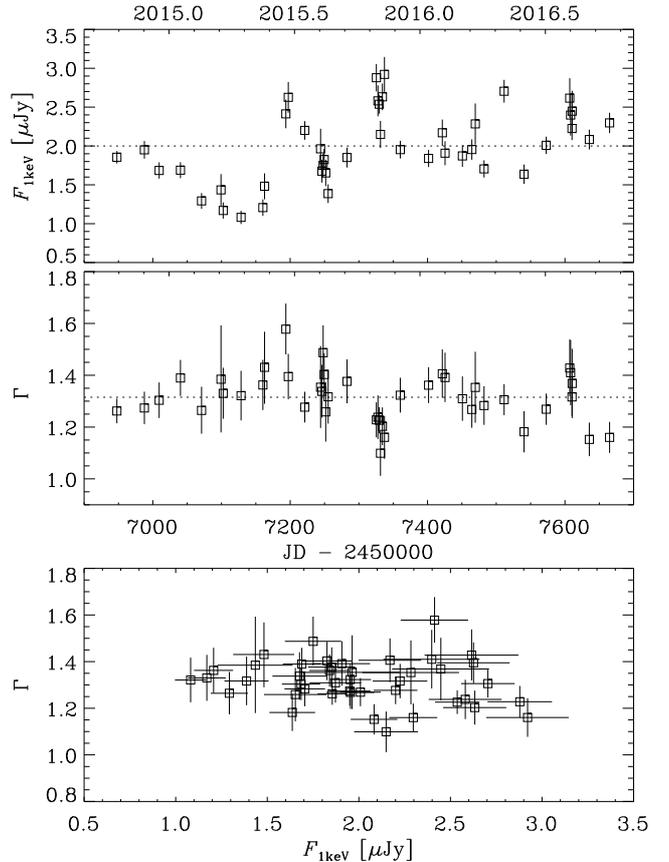}
    \caption{Results of the {\it Swift}-XRT data analysis when adopting a power-law model with absorption fixed to $N_{\rm H}^{\rm best}=6.3 \times 10^{20} \rm \, cm^{-2}$. Top: X-ray flux density at 1 keV versus time. Middle: photon index $\Gamma$ versus time. Bottom: photon index versus flux density. }
    \label{xrt_flux_gamma}
\end{figure}

We then performed a third fitting run, where the spectra are modelled with a power law with $N_{\rm H}$ fixed to the best-guess value. The corresponding $\chi^2/\nu$ are shown in Fig.\ \ref{xrt_chi2}. In general, they are very close to the values obtained in the power law with $N_{\rm H}$ free case. 
The $\rm 1 \, keV$ flux densities and photon indices $\Gamma$ are shown in Fig.\ \ref{xrt_flux_gamma}.
The 1 keV flux ranges from 1.08 to 2.92 $\mu$Jy, with a mean value of 2.00 $\mu$Jy and standard deviation of 0.47 $\mu$Jy, while $\Gamma$ goes from 1.10 to 1.58, with a mean value of 1.32 and standard deviation of 0.09. This case implies slighly softer spectra than in the case where $N_{\rm H}$ is fixed to the Galactic value, which yields a mean photon index of 1.23. We note that there is no correlation between the flux and the photon index.

We finally tested the effects of a spectral curvature by fitting a log-parabola model with absorption fixed to $N_{\rm H}^{\rm best}$. The curvature parameter showed a very large spread with large uncertainties, so we believe that this case cannot add meaningful information.

\section{Observations by {\it Fermi}}

We processed the Pass 8 data \citep{atw13} from the Large Area Telescope \citep[LAT;][]{atw09} onboard the \textit{Fermi} satellite using the {\sevensize \bf SCIENCETOOLS} software package version v10r0p5 and following standard procedures \citep[see e.g.][]{carnerero2015}.

\begin{figure}
	\includegraphics[width=\columnwidth]{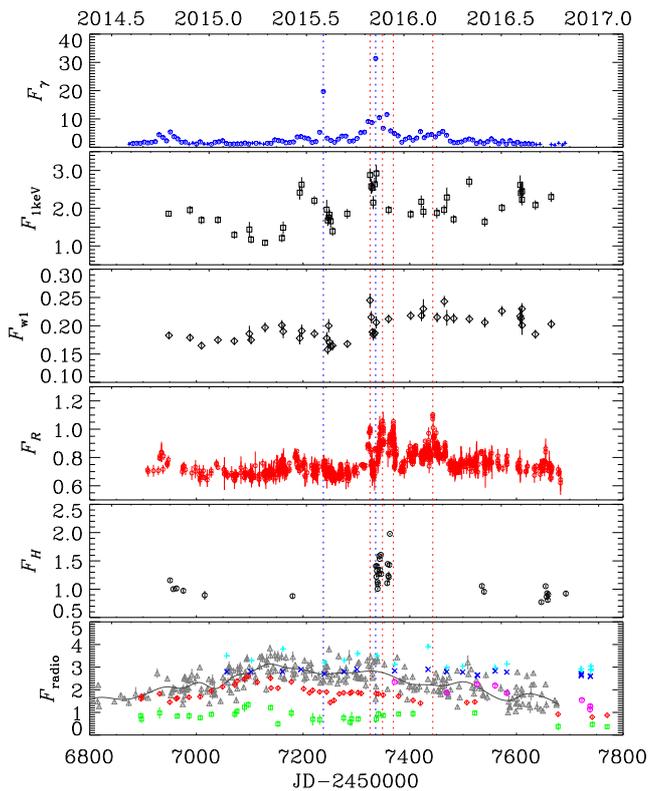}
    \caption{Multiwavelength light curves of 4C 71.07. From top to bottom: 
{\it i)} 0.1--300 GeV fluxes (circles, $10^{-7} \rm \, ph \, cm^{-2} \, s^{-1}$) and upper limits (plus signs) from {\it Fermi}-LAT; 
{\it ii)} 1 keV flux densities ($\mu \rm Jy$) from {\it Swift}-XRT;
{\it iii)} {\it Swift}-UVOT flux densities (mJy) in $w$1 band corrected for the Galactic, but not for IGM absorption;
{\it iv)} flux densities (mJy) in $R$ band corrected for the Galactic extinction;
{\it v)} flux densities (mJy) in $H$ band corrected for the Galactic extinction;
{\it vi)} radio flux densities (Jy) at 5 GHz (cyan plus signs), 8 GHz (blue crosses), 24 GHz (magenta circles),
37 GHz (grey triangles; the solid line represents a cubic spline interpolation on the 30-day binned data), 86 GHz (red diamonds), 228 GHz (green squares).
The blue and red vertical lines guide the eye through the $\gamma$-ray and optical peaks, respectively.
} 
    \label{mw}
\end{figure}

We considered both a power-law and a log-parabola model for the source spectrum and a week time bin for the light curve. In Fig.\ \ref{mw} we show the $\gamma$-ray light curve in the 0.1--300 GeV energy range resulting from the power-law fit.

\section{Broad-band multiwavelength behaviour}

Figure \ref{mw} compares the behaviour of 4C~71.07 at different wavelengths, from the $\gamma$ rays to the radio band.
The $\gamma$-ray light curve shows two prominent maxima at $\rm JD=2457238$ and $\rm JD=2457336$. We lack {\it Swift} observations at the time of the first maximum, but the X-ray flux was in a high state shortly before. The X-ray data exhibit a peak simultaneous to the second $\gamma$-ray maximum. Interestingly, another X-ray peak at $\rm JD=2457325$ corresponds to the maximum of the ultraviolet light curve, which in turn correlates with a major optical flare. The following three optical maxima are not covered by {\it Swift} observations. Their simultaneous $\gamma$-ray data show moderately high states.
Other remarkable X-ray levels are reached at JD=2457197, 2457511 and 2457607. The first event was preceeded by a minor optical flare and a mild flux increase is also visible in the $\gamma$-rays. The second event has a possible, weak $\gamma$-ray, but no optical, counterpart, and the third event does not seem to have any counterpart at all.

The near-infrared light curves are undersampled, so we can just notice higher fluxes in the period of increased optical activity. As for the radio bands, the 5 GHz data seem rather scattered, those at 8 GHz show little variability, while at 37, 86, and 228 GHz the flux maximum is reached before the optical, X-ray and $\gamma$-ray activity and then the radio flux declines.

\begin{figure}
	\includegraphics[width=\columnwidth]{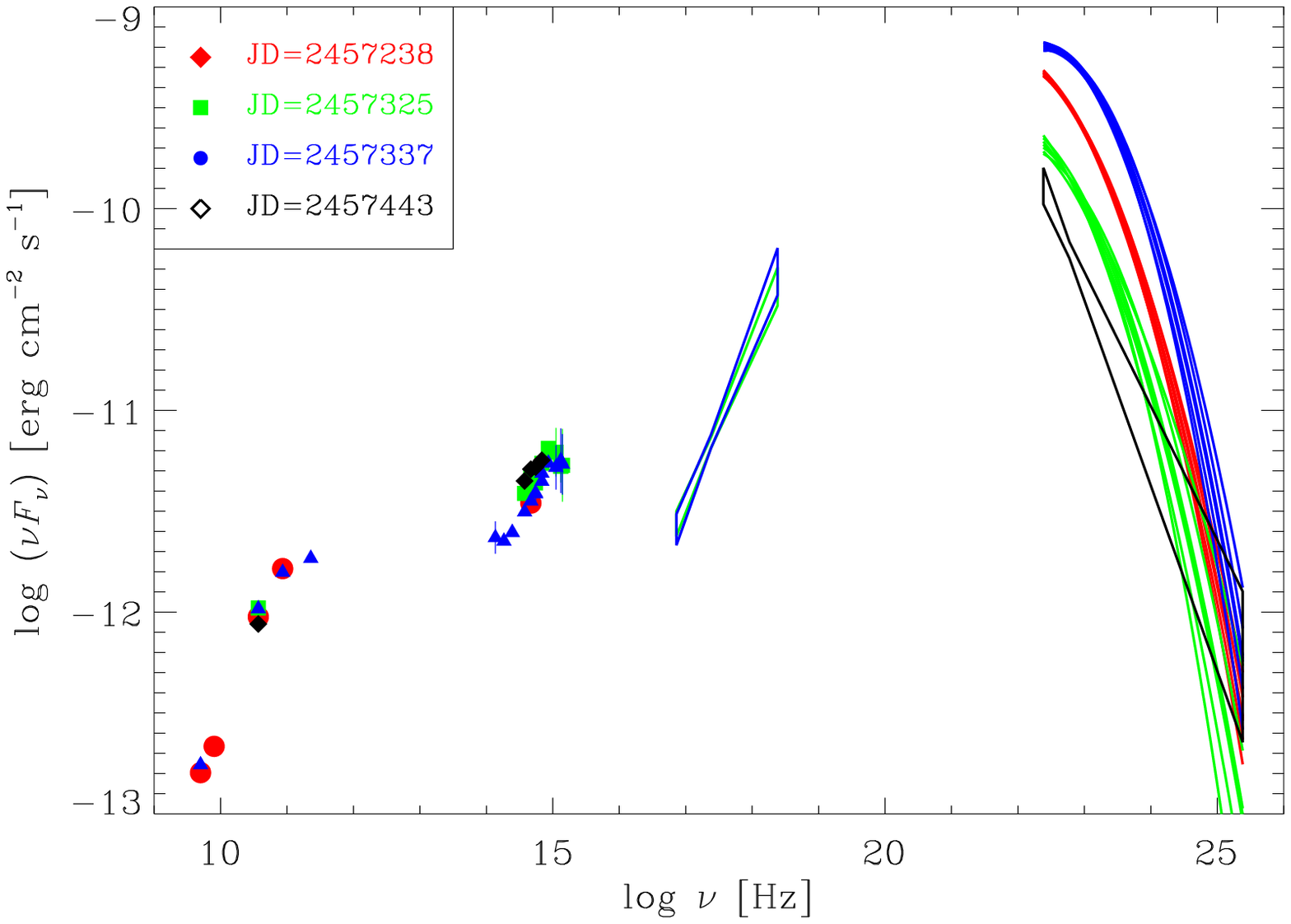}
 	\includegraphics[width=\columnwidth]{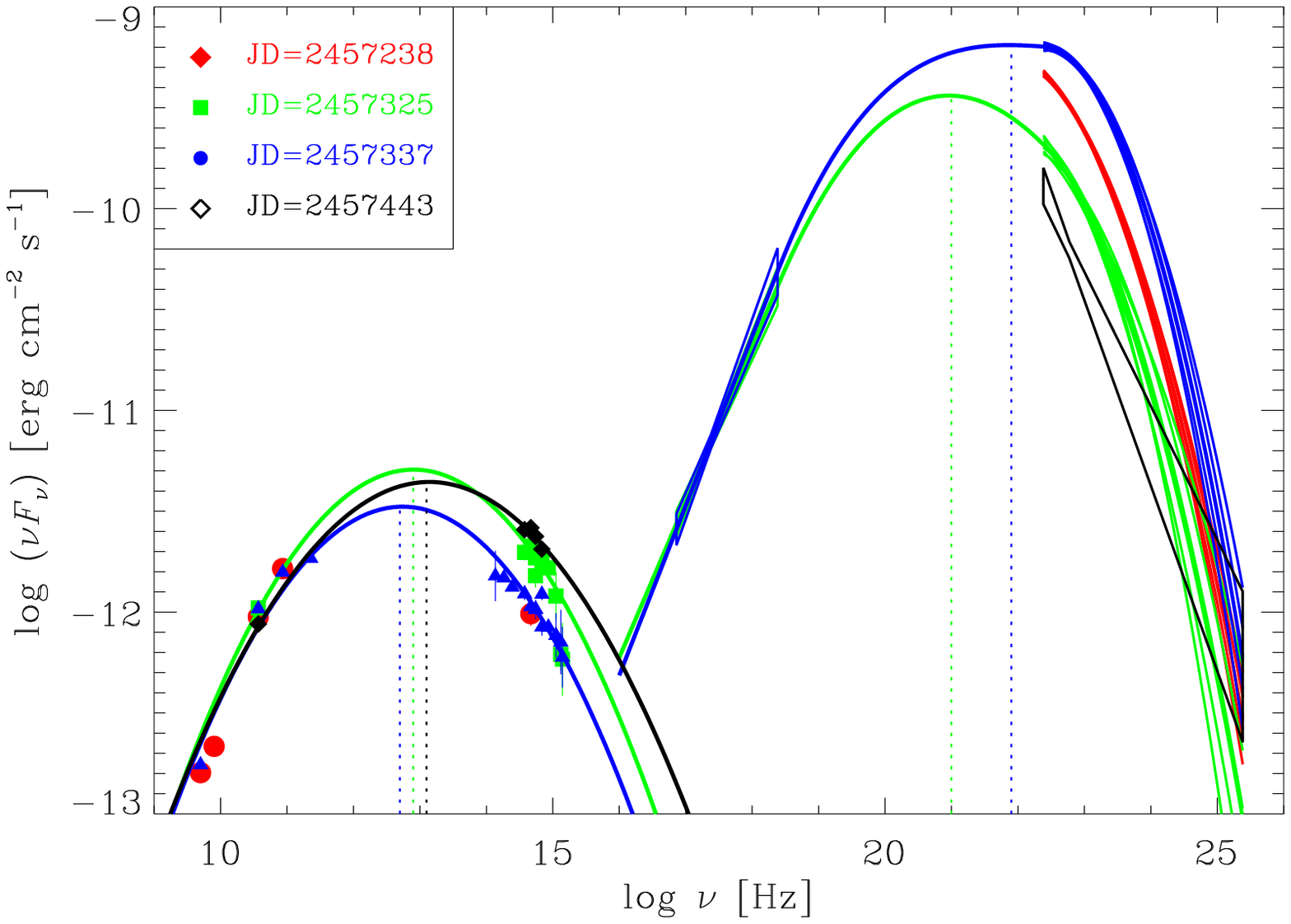}
    \caption{Broad-band SEDs of 4C~71.07 at four epochs, indicated in the upper left inset. Near-infrared and optical data have been corrected for the Galactic extinction. In the top panel, data in the $u$ and ultraviolet bands have also been corrected for the IGM absorption, so that these SEDs represent the total 
deabsorbed source emission. In the bottom panel, the big blue bump contributions estimated from the template shown in Fig.\ \ref{sed_template} have been subtracted from the near-infrared, optical and ultraviolet flux densities, so that these SEDs represent the synchrotron jet emission. The X-ray spectra have been obtained with a power-law model with absorption fixed to our best-guess value $N_{\rm H}^{\rm best}=6.3 \times 10^{20} \rm \, cm^{-2}$. $\gamma$-ray spectra have been modelled with a log-parabola except for the faintest state, where a power law was preferred. In the bottom panel the solid lines are log-parabolic fits and cubic spline interpolations to the synchrotron and inverse-Compton components, respectively, to highlight the possible shift of the bump peaks toward higher energies with increasing flux.}
   \label{broad}
\end{figure}

Figure \ref{broad} shows broad-band SEDs built with data simultaneous at all frequencies but in the radio band, where a few days distance was accepted because of the smoother flux variations at those wavelengths.
We chose four epochs, corresponding to different $\gamma$-ray and optical brightness states. Two SEDs correspond to the two peaks in the $\gamma$-ray light curve, while the other two SEDs refer to the first and the fourth optical maxima.
{\it Swift} data are available for only two SEDs. The plotted XRT spectra are obtained with a power-law model with absorption fixed to our best-guess value $N_{\rm H}^{\rm best}=6.3 \times 10^{20} \rm \, cm^{-2}$. {\it Fermi}-LAT spectra are fitted with log-parabolic models but in the faintest state, where we show the results of a power-law fit, as the curvature parameter of the log-parabolic model was very small.

In the upper panel the data in the $u$ and ultraviolet bands have been corrected for IGM absorption.
Therefore, these SEDs represent the total deabsorbed emission of the source, including the quasar core and beamed jet contributions.

In the bottom panel, we subtracted the big blue bump contributions from the near-infrared, optical and ultraviolet fluxes according to the prescriptions given in Section \ref{template}. These SEDs thus represent the pure jet emission.
In this representation it is easier to see that increasing brightness states in the near-infrared to ultraviolet bands correspond to decreasing brightness states in $\gamma$ rays.

Log-parabolic fits to the data from the radio to the ultraviolet band highlight that the frequency of the synchrotron peak shifts toward higher values as the near-infrared-to-ultraviolet spectrum rises.
A simple interpolation through the X-ray and $\gamma$-ray data allows us to verify that the same shift also applies to the inverse-Compton peak. Moreover, we can give a rough estimate of the Compton dominance, i.e.\ the ratio between $\nu \, F_\nu$ of the two peaks, in the two epochs where {\it Fermi} data are available. This ratio is about 70 at $\rm JD=2457325$ and about 200 at $\rm JD=2457337$.

The jet emission SEDs allow us to estimate the jet bolometric luminosity by integrating the synchrotron and inverse-Compton bumps. The result for the $\rm JD=2457325$ epoch, which represents a somewhat mean state, is $L_{\rm jet}=9.42 \times 10^{49} \rm \, erg \, s^{-1}$, 98.5\% of which is due to the high-energy bump. This extremely high luminosity is linked to the jet power spent in radiation by the relation \citep{ghisellini2014}: $P_{\rm rad} \simeq 2 \, L_{\rm jet} / \Gamma^2$, where $\Gamma$ is the bulk Lorentz factor. By considering a range of possible $\Gamma$ values, from $\Gamma=14$ \citep{ghisellini2010} to $\Gamma=28$ \citep{savolainen2010}, we find $P_{\rm rad}=(2.40$--$9.61) \times 10^{47} {\rm \, erg \, s^{-1}} \simeq (1$--$4) \, L_{\rm disc}$. 
This puts 4C~71.07 close (within 1--$2 \, \sigma$) to the best-fit linear correlation between radiative jet power and disc luminosity derived by \citet{ghisellini2014} and confirms the validity of this relation at the highest blazar energies.

\section{Polarimetric observations}

Synchrotron emission is polarised and in blazars the degree of polarisation ($P$) and electric vector polarisation angle (EVPA) can be extremely variable \citep[e.g.][]{smith1996}. The polarisation properties are expected to mirror the properties of the magnetic field in the emission region(s) and hence they can potentially shed light on the jet physics and structure. Actually, it is not clear yet to what extent stochastic processes due to turbulence act in determining the polarisation behaviour in blazars \citep[e.g.][]{marscher2014,kiehlmann2017,raiteri2017}. Large EVPA rotations have been observed that are sometimes correlated with flares detected at $\gamma$-rays \citep[e.g.][]{blinov2018}. The picture appears quite complex, as changes in the jet viewing angle can mimic a stochastic process even when the variations in flux and $P$, and EVPA rotations, are produced by a deterministic process \citep{lyutikov2017}.

\begin{figure}
	\includegraphics[width=\columnwidth]{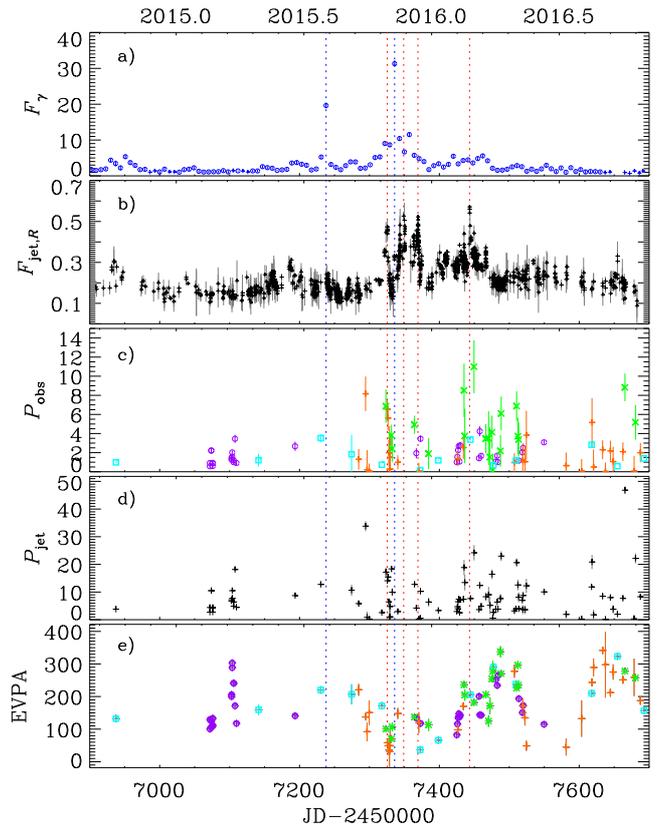}
    \caption{From top to bottom: a) 0.1--300 GeV $\gamma$-ray flux ($10^{-7} \rm \, ph \, cm^{-2} \, s^{-1}$) from {\it Fermi}-LAT; b) $R$-band deabsorbed jet (black plus signs) flux densities (mJy); c) observed degree of polarisation (\%); data are from Calar Alto (cyan squares), Crimean (orange plus signs), Lowell (violet circles), and St.\ Petersburg (green crosses) observatories; d) degree of polarisation (\%) of the jet contribution (black squares), after correcting the observed values for the dilution effect due to the big blue bump; e) EVPA (degree; symbols and colours as in panel c) adjusted for the $\pm n \times \pi$ ambiguity. Blue and red vertical dotted lines are drawn to guide the eye through the $\gamma$-ray and optical major peaks respectively.}
    \label{pola}
\end{figure}

Polarisation data for this work were acquired at the Calar Alto, Crimean, Lowell, and St.\ Petersburg observatories. The $P$ and EVPA behaviour in time is plotted in Fig.\ \ref{pola} and compared with the $\gamma$ and optical light curves.
The optical light curve shows the contribution of the jet to the $R$-band flux densities,
$F_{\rm jet}=F_{\rm tot}-F_{\rm BBB}$, where $F_{\rm BBB}=0.532 \rm \, mJy$ is the big blue bump contribution, as obtained in Sect.\ \ref{template} and listed in Table \ref{dati}. 

\begin{figure}
	\includegraphics[width=\columnwidth]{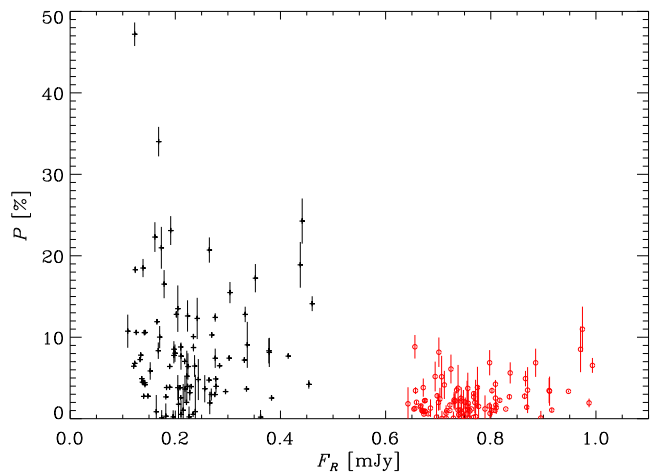}
    \caption{Degree of polarisation versus deabsorbed flux density. Black plus signs refer to the jet component, while red circles to observed $P$ and total flux density.}
    \label{pf}
\end{figure}

The observed degree of polarisation varies very fast, ranging from about zero (0.03\%) to 11\%. We have no data simultaneous with the $\gamma$-ray and optical peaks, except for the first major optical peak at $\rm JD=2457325.5$, where $P_{\rm obs}=6.5 \%$. The somewhat sparse sampling and large errors affecting many polarisation data points prevent us to draw firm conclusions, but there seems to be a lack of correlation between $P$ and the flux (see also Fig.\ \ref{pf}) that remains true even when we correct $P_{\rm obs}$ for the dilution effect of the big blue bump to derive the degree of polarisation of the jet:
$P_{\rm jet}= (F_{\rm tot} \times P_{\rm obs})/F_{\rm jet}.$
The minimum and maximum values of $P_{\rm jet}$ are 0.11\% and 47\%, respectively. The maximum is more than four times higher than the observed maximum value.
We recall that the degree of polarisation expected for synchrotron radiation from a power-law distribution of particles is $P_{\rm syn} = (p+1)/(p+7/3)= 0.69-0.75$ for typical power-law indices $p = 2-3$ \citep{rybicki1979}. 
A value for $F_{\rm BBB}$ higher than what we have assumed would further increase $P_{\rm jet}$, pushing its maximum toward the above theoretical value. Therefore, polarisation can potentially be used to constrain the emission contribution from the big blue bump in FSRQs.

The EVPA presents a $\pm n \times \pi$ ambiguity that can be treated giving a reasonable prescription. The values shown in Fig.\ \ref{pola} have been obtained by simply minimising the difference between subsequent angles. EVPA rotations both clockwise and counter-clockwise are recognizable, a behaviour that has often been ascribed to a stochastic process due to turbulence \citep{marscher2014,raiteri2017}.

A fast and large clockwise rotation of about 180\degr\ occurs around $\rm JD \sim 2457104$--10, during a period of low $\gamma$ and optical activity. A counter-clockwise rotation of $\sim 140$\degr\ in 12 days preceeds the optical ``sterile" flare (i.e.\ withouth $\gamma$-ray counterpart) at $\rm JD=2457443$. Other remarkable rotations are seen starting at $\rm JD=2457470$ (166\degr\ in a week, counter-clockwise) and at $\rm JD=2457507$ (143\degr\ in 15 days and possibly 230\degr\ in 17 days, clockwise), in periods when no significant optical or $\gamma$-ray flux variations are observed.

\section{Conclusions}

Because of the beamed nature of the source, the blazar emission that we observe is usually dominated by the non-thermal radiation from the jet. However, FSRQs often receive considerable thermal emission contributions from their quasar cores, which peak at wavelengths increasing with redshift. But as the redshift becomes higher, the more is the intergalactic medium that the blazar radiation must cross and hence the stronger the absorption. The problem of disentangling the jet and big blue bump contributions to the observed emission and that of correcting the observed fluxes for the IGM absorption are key issues when investigating the properties of high-redshift FSRQs.

In this paper we have analysed the flux and polarimetric behaviour of the high-redshift FSRQ-type blazar 4C~71.07. Optical, near-infrared and radio light curves have been built with data taken by the WEBT during an intensive campaign in 2014--2016. These have been complemented by ultraviolet and X-ray data from {\it Swift} and by $\gamma$-ray data by {\it Fermi}. 

During the campaign we also obtained optical spectroscopic data that have been analysed in detail by \citet{raiteri2019}. We use here the average optical and near-infrared spectra to construct a model for the source quasar core. We filled the gaps between the near-infrared and the optical frequency range and in the ultraviolet by making use of the quasar templates by \citet{polletta2007} and \citet{lusso2015}, respectively. From the model we derived the flux contributions of the big blue bump to the source photometry in the various near-infrared, optical, and ultraviolet bands. These can be subtracted from the observed fluxes to obtain the synchrotron jet fluxes.
Following \citet{ghisellini2010} and \citet{lusso2015}, we also estimate the opacity values that can be applied to the UVOT data to correct for the IGM absorption.

We analysed the XRT data with different models and estimated a best-guess value for the total absorption due to both the Galaxy and IGM of $N_{\rm H}^{\rm best}=6.3 \times 10^{20} \rm \, cm^{-2}$, which is more than twice the Galactic value. The modest number of counts of the XRT spectra does not allow us to distinguish whether some intrinsic spectral curvature is present.

Light curves at different frequencies do not show persistent correlation, in particular among $\gamma$-rays, X-rays and optical fluxes.

Broad-band SEDs present, beside a prominent big blue bump, a very strong Compton dominance. The correction for the IGM absorption makes the ultraviolet spectrum harder and the X-ray spectrum softer, and this implies a smoother connection between them.

We verified that 4C~71.07 is characterised by extreme nuclear and jet properties. Integration of the thermal continuum traced by our big blue bump template allowed us to estimate the disc bolometric luminosity, $L_{\rm disc}=2.45 \times 10^{47} \rm \, erg \, s^{-1}$, and to derive the mass accretion rate, $\dot{M} \simeq 18 \rm \, M_\odot \, yr^{-1}$, and black hole mass, $M_{\rm BH} \simeq 1.6 \times 10^9 \, \rm M_\odot$, from it in the case of a Shakura \& Sunyaev disc. As a consequence, the Eddington ratio is as high as 0.66. On the other hand, we could estimate the jet bolometric luminosity integrating the nuclear-subtracted SED, obtaining $L_{\rm jet}=9.42 \times 10^{49} \rm \, erg \, s^{-1}$. From this we calculated the jet radiation power, $P_{\rm rad}=(2.40$--$9.61) \times 10^{47} {\rm \, erg \, s^{-1}} \simeq (1$--$4) \, L_{\rm disc}$. The disc and jet luminosities of 4C~71.07 are thus found to fit fairly well into the jet-disc relation for blazars \citep{ghisellini2014}, confirming it at the highest energy values.

The polarisation data acquired during the WEBT campaign display strong variability in both the polarisation degree and polarisation angle. This variability seems not to be correlated with the flux behaviour. Correction of $P$ for the dilution effect of the thermal radiation from the quasar core brings the maximum from $\sim 11\%$ to $\sim 47\%$, but does not lead to a correlation with the flux. Noticeable EVPA rotations are observed, both clockwise and counter-clockwise. They mostly occur during  periods where the flux does not show significant changes and may be caused by turbulence.

In the light of our results, we conclude by stressing the importance of taking in due consideration the contribution of the quasar core when analysing the emission from FSRQs and the role of IGM absorption when dealing with high-redshift objects.

\section*{Acknowledgements}
We thank an anonymous referee for stimulating comments and Elisabeta Lusso and Gabriele Ghisellini for sharing results of previous publications.
The data collected by the WEBT collaboration are stored in the WEBT archive at the Osservatorio Astrofisico di Torino - INAF (http://www.oato.inaf.it/blazars/webt/); for questions regarding their availability, please contact the WEBT President Massimo Villata ({\tt massimo.villata@inaf.it}).
This publication uses data obtained at Mets\"ahovi Radio Observatory, operated by Aalto University in Finland.
We acknowledge financial contribution from the agreement ASI-INAF n.2017-14-H.0 and from the contract PRIN-SKA-CTA-INAF 2016. P.R. and S.V. acknowledge contract ASI-INAF I/004/11/0. 
We acknowledge support by Bulgarian National Science Programme ``Young Scientists and Postdoctoral Students 2019", Bulgarian National Science Fund under grant DN18-10/2017 and National RI Roadmap Projects DO1-157/28.08.2018 and DO1-153/28.08.2018 of the Ministry of Education and Science of the Republic of Bulgaria.
GD and OV gratefully acknowledge the observing grant support from the Institute of Astronomy and Rozhen National Astronomical Observatory, Bulgarian Academy of Sciences via bilateral joint research project ``Study of ICRF radio-sources and fast variable astronomical objects" (head - G.Damljanovic).
This work is a part of the Projects No. 176011 (``Dynamics and Kinematics of Celestial Bodies and Systems"), No. 176004 (``Stellar Physics") and No. 176021 (``Visible and Invisible Matter in Nearby Galaxies: Theory and Observations") supported by the Ministry of Education, Science and Technological Development of the Republic of Serbia. 
This research was partially supported by the Bulgarian National Science Fund of the Ministry of Education and Science under grants DN 08-1/2016, DN 18-13/2017 and KP-06-H28/3 (2018). 
The Skinakas Observatory is a collaborative project of the University of Crete, the Foundation for Research and Technology -- Hellas, and the Max-Planck-Institut  f\"ur Extraterrestrische Physik.
The St.Petersburg University team acknowledges support from Russian Science Foundation grant 17-12-01029.
The Abastumani team acknowledges financial support by the Shota Rustaveli National Science Foundation under contract FR/217950/16.
This work was partly supported by the National Science Fund of the Ministry of Education and Science of Bulgaria under grant DN 08-20/2016, and by funds of the project RD-08-37/2019 of the University of Shumen.
The Astronomical Observatory of the Autonomous Region of the Aosta Valley (OAVdA) is managed by the Fondazione Cl\'{e}ment Fillietroz-ONLUS, which is supported by the Regional Government of the Aosta Valley, the Town Municipality of Nus and the ``Unit\'{e} des Communes vald\^{o}taines Mont-\'{E}milius". The research at the OAVdA was partially funded by two ``Research and Education" grants from Fondazione CRT.
This research has made use of the NASA/IPAC Extragalactic Database (NED) which is operated by the Jet Propulsion Laboratory, California Institute of Technology, under contract with the National Aeronautics and Space Administration. 




\bibliographystyle{mnras}
\bibliography{/home/claudia/my} 


\vspace{1cm}\noindent
{\large \bf AFFILIATIONS}

\vspace{0.5cm}\noindent
{\it
$^{ 1}$INAF, Osservatorio Astrofisico di Torino, via Osservatorio 20, I-10025 Pino Torinese, Italy                                                           \\
$^{ 2}$Instituto de Astrofisica de Canarias (IAC), La Laguna, E-38200 Tenerife, Spain                                                                        \\
$^{ 3}$Departamento de Astrofisica, Universidad de La Laguna, La Laguna, E-38205 Tenerife, Spain                                                             \\
$^{ 4}$Ulugh Beg Astronomical Institute, Maidanak Observatory, Tashkent, 100052, Uzbekistan                                                                  \\
$^{ 5}$Astronomical Institute, St.\ Petersburg State University, 198504 St.\ Petersburg, Russia                                                              \\
$^{ 6}$Pulkovo Observatory, 196140 St.\ Petersburg, Russia                                                                                                   \\
$^{ 7}$INAF, Osservatorio Astronomico di Brera, Via Emilio Bianchi 46, 23807 Merate, LC, Italy                                                               \\
$^{ 8}$Instituto de Astrof\'{\i}sica de Andaluc\'{\i}a (CSIC), E-18080 Granada, Spain                                                                        \\
$^{ 9}$Max-Planck-Insitut f\"ur Radioastronomie, Auf dem H\"ugel 69, 53121 Bonn, Germany                                                                     \\
$^{10}$Institute of Astronomy and NAO, Bulgarian Academy of Sciences, 1784 Sofia, Bulgaria                                                                   \\
$^{11}$Dipartimento di Fisica G.\ Occhialini, Universit\`{a} degli Studi di Milano Bicocca, 20126 Milano, Italy                                                 \\
$^{12}$Crimean Astrophysical Observatory RAS, P/O Nauchny, 298409, Russia                                                                                    \\
$^{13}$INAF, TNG Fundaci\'on Galileo Galilei, E-38712 La Palma, Spain                                                                                        \\
$^{14}$Department of Astronomy, Faculty of Physics, University of Sofia, BG-1164 Sofia, Bulgaria                                                             \\
$^{15}$Osservatorio Astronomico della Regione Autonoma Valle d'Aosta, I-11020 Nus, Italy                                                                     \\
$^{16}$EPT Observatories, Tijarafe, E-38780 La Palma, Spain                                                                                                  \\
$^{17}$Graduate Institute of Astronomy, National Central University, Jhongli City, Taoyuan County 32001, Taiwan                                              \\
$^{18}$Astronomical Observatory, 11060 Belgrade, Serbia                                                                                                      \\
$^{19}$INAF, Osservatorio Astronomico di Roma, I-00040 Monte Porzio Catone, Italy                                                                            \\
$^{20}$Sternberg Astronomical Institute, M.V. Lomonosov Moscow State University, Moscow 119991, Russia                                                       \\
$^{21}$INAF, Istituto di Radioastronomia, Via Piero Gobetti 93/2, 40129 Bologna, Italy                                                                       \\
$^{22}$Department of Physics and Astronomy, Faculty of Natural Sciences, University of Shumen, 9700 Shumen, Bulgaria                                         \\
$^{23}$Astrophysics Research Institute, Liverpool John Moores University, Liverpool L3 5RF, UK                                                               \\
$^{24}$Institute for Astrophysical Research, Boston University, Boston, MA 02215, USA                                                                        \\
$^{25}$Abastumani Observatory, Mt. Kanobili, 0301 Abastumani, Georgia                                                                                        \\
$^{26}$Engelhardt Astronomical Observatory, Kazan Federal University, Tatarstan, Russia                                                                      \\
$^{27}$Aalto University Mets\"ahovi Radio Observatory, FI-02540 Kylm\"al\"a, Finland                                                                         \\
$^{28}$Aalto University Dept of Electronics and Nanoengineering, FI-00076 Aalto, Finland                                                                     \\
$^{29}$Department of Physics and Astronomy, Brigham Young University, Provo, UT 84602, USA                                                                   \\
$^{30}$Osservatorio Astronomico Sirio, I-70013 Castellana Grotte, Italy                                                                                      \\
$^{31}$Department of Physics, University of Colorado Denver, CO, 80217-3364 USA                                                                              \\
 }

\bsp	
\label{lastpage}
\end{document}